\documentclass[preprint,nofootinbib]{revtex4}
\usepackage{graphicx}

\def\gev{\rm GeV}
\def\mev{\rm MeV}
\def\ev{\rm eV}
\def\s{\rm s}
\def\nul{\nu_L^{}}
\def\nur{\nu_R^{}}
\def\dl{\Delta L}
\def\pslash{\not{\hbox{\kern-4pt $p$}}}
\def\qslash{\not{\hbox{\kern-4pt $q$}}}
\def\lv{\not{\hbox{\kern-4pt $L$}}}

\def\lsim{\mathrel{\raise.3ex\hbox{$<$\kern-.75em\lower1ex\hbox{$\sim$}}}}
\def\gsim{\mathrel{\raise.3ex\hbox{$>$\kern-.75em\lower1ex\hbox{$\sim$}}}}
\def\ifmath#1{\relax\ifmmode #1\else $#1$\fi}

\begin{document}

\bibliographystyle{revtex}

\preprint{
 {\vbox{
 \hbox{\bf MADPH--04--1393}
  \hbox{\bf FERMILAB-PUB-04-336-T  }
    \hbox{\bf ANL-HEP-PR-05-7 }
 \hbox{hep-ph/0502163}}}}

\vspace*{2cm}

\title{Upper Bounds on Lepton-number Violating Processes}

\author{Anupama Atre$^1$, Vernon Barger$^1$,
 Tao Han$^{1,2,3}$\footnote{avatre@physics.wisc.edu,\\ 
barger@physics.wisc.edu,\\ than@physics.wisc.edu }}
\affiliation{\vspace*{0.1in}
$^1$Department of Physics, University of Wisconsin, 1150 University
Avenue, Madison, WI 53706\\
        $^2$Theoretical Physics Department, Fermi National Accelerator Laboratory, 
        P.O.Box 500, MS106, Batavia, IL 60510\\
                $^3$Theory Group, High Energy Physics Division,
                Argonne National Laboratory,  Argonne, IL 60439
\vspace*{.25in}}  %% makes space between address and abstract 

\begin{abstract}
%\noindent
We consider four lepton-number violating ($\lv$) processes: 
(a) neutrinoless double-beta decay $(0\nu\beta\beta)$, (b) $\dl=2$ tau decays, 
(c) $\dl=2$ rare meson decays and (d) nuclear muon-positron conversion. 
In the absence of exotic $\lv$ interactions, the rates for these processes 
are determined by effective neutrino masses $\left<m\right>_{\ell_1\ell_2}$, 
which can be related to the sum of light neutrino masses, the neutrino mass-squared differences, 
the neutrino mixing angles, a Dirac phase and two Majorana phases. 
We sample the experimentally allowed ranges of $\left<m\right>_{\ell_1\ell_2}$ 
based on neutrino oscillation experiments as well as cosmological
observations, and obtain a stringent upper bound 
$\left<m\right>_{\ell_1\ell_2}\lsim 0.14$ eV. 
We then calculate the allowed ranges for $\left<m\right>_{\ell_1\ell_2}$ from the experimental rates of direct searches for the above $\dl=2$ processes. Comparing our calculated rates with
the currently  or soon available data, we find that only the $0\nu\beta\beta$ 
experiment may be able to probe $\left<m\right>_{ee}$ with a sensitivity
comparable to the current bound. Muon-positron conversion is next in sensitivity,  while the limits of direct searches for the other $\dl=2$
processes are several orders of magnitude weaker  than the current bounds  on 
$\left<m\right>_{\ell_1\ell_2}$. Any positive signal in those direct searches would
indicate new contributions to the $\lv$ interactions beyond those from three light Majorana
neutrinos.
\end{abstract}

\maketitle

\section{Introduction}

Fermion masses and flavor mixing are among the most mysterious problems of 
contemporary particle physics and they have posed major challenges to 
particle theory and experiment for decades. 
Further understanding of these issues should eventually 
shed light on fundamental phenomena like CP violation, 
flavor-changing neutral currents, 
baryon-number ($B$) and lepton-number ($L$) asymmetry in the Universe
and will hopefully lead to a more satisfactory unified theory of flavor physics \cite{review}.
 
In the Standard Model (SM) of strong and electroweak interactions, neutrinos
are strictly massless due to the absence of the right-handed chiral states 
($\nu_R^{}$) and the requirement of $SU(2)_L$ gauge invariance and 
renormalizability. Recent neutrino oscillation 
experiments have conclusively shown that  neutrinos are 
massive \cite{BargerReview}.  This discovery presents a pressing need to
consider physics beyond the Standard Model. It is straightforward to introduce 
a Dirac mass term   $m_D^{} (\overline{\nul} \nur + $h.c.) for a neutrino 
by including the right-handed state, just like the treatment for all other fermions
via the Yukawa couplings to the Higgs doublet 
in the SM. However,  a profound question arises:  Since $\nur$ is a SM gauge
singlet, why should not there  exist  a gauge-invariant 
Majorana mass term ${1\over 2}M \nur \nur$ in the  theory? 
In fact, there is a strong theoretical motivation for the Majorana mass term 
to exist since it could naturally explain the smallness of the observed
neutrino masses via the so-called ``see-saw"  mechanism 
$m_\nu \approx m^2_D/M$ \cite{seesaw}.  From a model-building point of view,
there are many scenarios that could incorporate the Majorana mass. Examples
include R-parity violating interactions ($\dl=1$) in Supersymmetry (SUSY) \cite{RparitySUSY},  
Left-Right symmetric gauge theories \cite{LRModels}, grand unified theories \cite{SO10SUSYGUT}, models  with exotic Higgs representations \cite{ZeeModel,MaModels} 
and theories with extra dimensions \cite{ExtraDim}. One may also consider constructing generic neutrino mass operators to parameterize the fundamental 
physics  effects in a model-independent  manner \cite{babuleung}.

\begin{figure}[tb]
\includegraphics[width=2in]{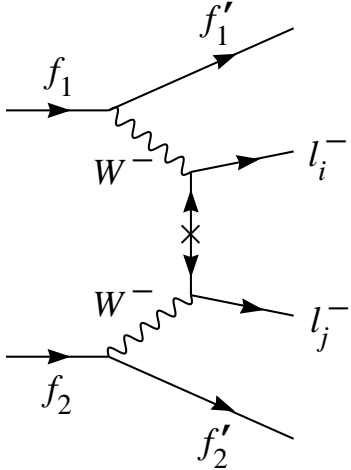}
\caption{A generic diagram for $\dl=2$ processes via Majorana neutrino exchange. }
\label{wwll}
\end{figure}

Besides the phenomena of neutrino flavor oscillations and possible new CP-violating
phases, the Majorana mass term violates 
 lepton number by two units ($\dl=2$), which may result in important consequences
in particle physics and cosmology. Although the prevailing theoretical prejudice 
prefers Majorana neutrinos, experimentally 
testing the nature of the neutrinos, and lepton-number violation ($\lv$)
in general, is of fundamental importance.
The basic process with $\dl=2$ is mediated by
\begin{eqnarray}
W^-W^- \rightarrow \ell^-_1 \ell^-_2,
\nonumber
\end{eqnarray}
where the $W^-$ are virtual SM weak bosons and $\ell_{1,2} =e,\ \mu,\ \tau$. 
By coupling fermion currents to the $W$ bosons as depicted in Fig.~\ref{wwll}, and
arranging the initial and final states properly, one finds various physical processes
that can be experimentally searched for. The best known example is the 
neutrinoless double-beta decay ($0\nu\beta\beta$) \cite{nuless, ElliottEngel, DoiKotani}, which proceeds via the parton-level subprocess $dd \to uu\ W^{-*} W^{-*} \to uu\ e^- e^-$. Other 
interesting classes of $\lv$ processes involve tau decays such as
$\tau^- \to e^+ (\mu^+) \pi^- \pi^-$ etc. \cite{TauDecay} and hyperon decays 
such as $\Sigma^- \to \Sigma^+ e^- e^-$, $\Xi^- \to p \mu^- \mu^-$ etc. \cite{Barbero}.
One could also explore additional processes like $e^- \to \mu^+$ conversion \cite{CSLim} .
One may also consider searching for signals at  accelerator and collider experiments 
via $e^-e^- \to W^-W^-$ \cite{TRizzo}, 
$e^\pm p \to \nu_e(\bar \nu_e)\ell_1^\pm \ell_2^\pm X$ \cite{Rodejohann}, 
$\nu_\ell(\bar \nu_\ell) N \to \ell^\mp \ell_1^\pm \ell_2^\pm X$ \cite{Flanz} and $pp \to  \ell^+_1 \ell^+_2 X$ \cite{AliBorisov}.

Assuming no additional contributions from other exotic particles that
have $\lv$ interactions, the matrix element for $\lv$ processes is
proportional to the product of two flavor mixing matrix elements and
a $\lv$ mass insertion from a light Majorana neutrino
\begin{eqnarray}
\nonumber
\left<m\right>_{\ell_1\ell_2} = | \sum_i V_{\ell_1 i}V_{\ell_2 i} m_i | .
\end{eqnarray}
The $\left<m\right>_{\ell_1\ell_2}$ are called ``effective neutrino masses". 
Experimental searches for $\lv$ processes
will directly measure the effective neutrino masses squared, and thus probe
the fundamental parameters of neutrino mixing angles and phases and 
their masses.

In this paper we study  $\dl=2$  processes related to 
$W^-W^- \rightarrow \ell_1^-\ell_2^-$.
We establish our conventions and lay out the general expressions for the effective 
neutrino masses in Sec.~\ref{enm}. We then calculate those quantities, based on
the current knowledge from atmospheric, solar and reactor neutrino oscillation 
experiments. We include the constraints on
neutrino masses from  a joint analysis of Wilkinson Microwave Anisotropy Probe (WMAP), Cosmic Microwave Background (CMB) data, the Sloan Digital Sky Survey (SDSS) large scale galaxy 
survey and bias, 
the SDSS Ly$\alpha$ forest power spectrum and the latest supernovae SNIa sample.
We thereby  limit the allowed ranges of the $\left<m\right>_{\ell_1 \ell_2}$. 
In Sections \ref{ee0n}, \ref{taud}, \ref{mesond}, and \ref{mue} 
we study  the four $\dl=2$  processes, 
\begin{itemize}
\item[(a)]  neutrinoless double-beta decay ($0\nu\beta\beta$), 
\item[(b)]  $\lv$ tau decays $\tau^- \rightarrow \ell^+ M_1^- M_2^-$, 
\item[(c)]  $\lv$ meson decays $M_1^+ \rightarrow M_2^- \ell_1^+ \ell_2^+$,
\item[(d)]  nuclear muon-positron ($\mu^--e^+$) conversion.
\end{itemize}
We calculate the transition rates and determine the allowed ranges 
of $\left<m\right>_{\ell_1 \ell_2}$ from the bounds set by direct experimental searches 
for the above $\dl=2$ processes, and then compare the results with the bounds 
obtained based on the neutrino mixing inputs in Sec.~\ref{enm}. We find that the upper bound
from $0\nu\beta\beta$ is  most sensitive to the model-parameters
and is at the same level as the constraint from Sec.~\ref{enm}. 
The bounds from the other three classes of measurements are significantly weaker
although they probe different combinations  of the parameters.  
In the future should we 
observe a $\lv$ signal in one of those channels, it would indicate non-standard
physics beyond the contributions of light Majorana neutrinos.
We draw our conclusions in Sec.~\ref{concld}.
Some technical details in calculating the transition rates for the above processes 
are presented in the Appendices. 

%%%%%%%%%%%%%%%%%%%%%%%%%%%%%%%%%%%%%%%%%%%%%%%%%%%%%%%%%%%%%
\section{Effective Neutrino Masses}
\label{enm}
%%%%%%%%%%%%%%%%%%%%%%%%%%%%%%%%%%%%%%%%%%%%%%%%%%%%%%%%%%%%%%%

In terms of neutrino mass eigenstates $\nu_i$ the charged current interaction Lagrangian is
written as 
\begin{equation}
\label{leff}
{\cal L}_{cc} = -\frac{g}{\sqrt{2}} \sum_{\ell = e,\mu,\tau} {\sum_{i=1,2,3}}V_{\ell i} {\bar \ell} 
\gamma^\mu P_L \nu_i W_\mu + {\rm h.c.}
\end{equation}
where $P_L$ is the left-handed projection operator $(1 - \gamma_5)/2$, 
$V_{\ell i}$ is the Maki-Nakamura-Sakata-Pontecorvo (MNSP) mixing matrix 
element \cite{MNSP}  between the lepton mass eigenstate $\ell=e, \mu, \tau$ and
the $i^{th}$ neutrino mass eigenstate.
%$\nu_\ell = \sum\limits_{i=1,2,3} {V^*_{\ell i}} {\nu_i}$. 
It is conveniently parameterized by \cite{BargerReview}
\begin{eqnarray}
\label{VMNSMatrix}
V = 
\left( \matrix{c_s c_x & s_s c_x e^{i\frac{\phi_2}{2}} & s_x e^{i\frac{\phi_3}{2}} \cr
-s_s c_a - s_a s_x c_s e^{i\delta}& c_a c_s e^{i\frac{\phi_2}{2}} - s_a s_s s_x e^{i\frac{\phi_2}{2}}e^{i\delta}  & s_a c_x e^{i\frac{\phi_3}{2}}e^{i\delta} \cr
s_a s_s - s_x c_a c_s e^{i\delta}  & -s_a c_s e^{i\frac{\phi_2}{2}} - s_s s_x c_a e^{i\frac{\phi_2}{2}}e^{i\delta}  & c_a c_x  e^{i\frac{\phi_3}{2}}e^{i\delta}  \cr} \right)
\end{eqnarray}
in the notation $c_i$ = cos$\theta_i$ and $s_i$ = sin$\theta_i$. The mixing angles $\theta_a$ and $\theta_s$ are relevant to atmospheric and solar oscillations, respectively, and the angle $\theta_x$ is presently unknown, except that it is bounded by the CHOOZ reactor data.
The phase $\delta$ is a Dirac phase and $\phi_2$, $\phi_3$ are Majorana phases.

The general subprocess of $\lv$ is neutrinoless dilepton production from 
two virtual $W$ bosons as depicted in Fig.~\ref{wwll}
\begin{equation}
W^-W^- \rightarrow \ell^-_1 \ell^-_2,
\label{basic}
\end{equation}
which  can occur only if neutrinos are Majorana particles. This subprocess 
changes the lepton number from $L=0$  to $L=2$ and the observation of 
$\lv$  would establish that  neutrino is a Majorana particle. 
The leptonic subprocess of Eq.~(\ref{basic}) occurs via Majorana neutrino exchange 
and is given by the product of  two charged  currents
\begin{equation}
\label{ampl}
{{\cal M}_{lep}^{\mu\nu}} \propto {\sum_i}V_{\ell_1i}V_{\ell_2i}({\bar \ell_1}
 \gamma^\mu P_L \nu_i)({\bar \ell_2} \gamma^\nu P_L \nu_i).
\end{equation}
As presented in Appendix \ref{appA}, 
the transition rates for light neutrino exchange 
 are proportional to the squares of effective neutrino masses 
 $\left<m\right>_{\ell_1\ell_2}$ defined as
\begin{equation}
\label{EffNuMass1}
\left<m\right>_{\ell_1 \ell_2} = {   \mid} \sum_i V_{\ell_1 i}V_{\ell_2 i} m_i \mid ,
\end{equation}
where $\ell_1,\ell_2 = e, \mu, \tau,\ i=1,2,3$. Since the mixing factor is symmetric 
$V_{\ell_1 i}V_{\ell_2 i} = V_{\ell_2 i}V_{\ell_1 i}$, 
there are six different $\left<m\right>_{\ell_1 \ell_2}$. 
The explicit expressions for  $\left<m\right>_{\ell_1 \ell_2}$ are given  \cite{Smirnov} by
\begin{eqnarray}
\label{Allmlls}
\nonumber
\left<m\right>_{ee} &=& |m_1 c_s^2  c_x^2 + m_2 s_s^2  c_x^2 e^{i\phi_2} + m_3 s_x^2 e^{i\phi_3}|, \\
\nonumber
\left<m\right>_{e \mu} &=& |m_1 c_s c_x (-s_s c_a - s_a s_x c_s e^{i\delta}) + m_2 s_s c_x (c_a c_s e^{i\phi_2} - s_a s_s s_x e^{i(\phi_2 + \delta)}) +  m_3 s_a s_x c_x e^{i(\phi_3 + \delta)}|,\\ 
\nonumber
\left<m\right>_{e\tau}& =& |m_1 c_s c_x (s_a s_s - s_x c_a c_s e^{i\delta}) + m_2 s_s c_x (-s_a c_s e^{i\phi_2} - s_s s_x c_a e^{i(\phi_2 + \delta)}) + m_3 s_x c_a c_x e^{i(\phi_3 + \delta)}|, \\
\nonumber
\left<m\right>_{\mu\mu}& =& |m_1 (s_s c_a +  s_a s_x c_s e^{i\delta})^2 + m_2 (c_a c_s e^{i\frac{\phi_2}{2}} - s_a s_s s_x e^{i(\frac{\phi_2}{2} + \delta)})^2 + m_3 s_a^2 c_x^2 e^{i(\phi_3 + 2\delta)}|,  \\
\nonumber
\left<m\right>_{\mu\tau}& =& |m_1 (s_s  c_a + s_a s_x c_s e^{i\delta})(s_x c_a c_s e^{i\delta} - s_a  s_s) + m_3 s_a  c_a c_x^2 e^{i(\phi_3 + 2\delta)}\\
\nonumber
 &+& m_2 (s_a s_s s_x e^{i(\phi_2 + \delta)} - c_a c_s e^{i\phi_2}) (s_a c_s + s_s s_x c_a e^{i\delta})|,  \\
\left<m\right>_{\tau\tau}& =& |m_1 (s_a s_s - s_x c_a c_s e^{i\delta})^2 + m_2(- s_a c_s  e^{i\frac{\phi_2}{2}} - s_s s_x c_a e^{i(\frac{\phi_2}{2} + \delta)})^2 + m_3 c_a^2 c_x^2 e^{i(\phi_3 + 2\delta)}|. 
\end{eqnarray}
We can see that the $\left<m\right>_{\ell_1 \ell_2}$ are functions of the oscillation angles ($\theta_a, \theta_s \mbox { and } \theta_x$) and the neutrino masses $m_i$. The three neutrino masses 
can be expressed in terms of other three  measured quantities:
the sum of neutrino masses and the two mass-squared differences, 
\begin{eqnarray}
\Sigma = m_1 + m_2 + m_3 = m_1 + \sqrt{m_1^2 + \delta m^2_s} + \sqrt{m_1^2 + \delta m^2_a} .
\label{SUM}
\end{eqnarray}
The atmospheric ($a$) and solar ($s$) mass-squared differences are defined as
\begin{eqnarray}
\label{MsqDiff}
\nonumber
\label{11a} 
\delta m^2_a &=& m^2_3 - m^2_1 ,   \\  
\delta m^2_s &=& m^2_2 - m^2_1 , 
\end{eqnarray}
 where 
 $\delta m^2_a > 0$ for the normal hierarchy (NH) and  $\delta m^2_a < 0$ for the 
 inverted hierarchy (IH).

The above expressions provide a convenient formalism 
to study the range of values of $\left<m\right>_{\ell_1 \ell_2}$ as functions of the angles and the phases (both Dirac and Majorana). 
In particular, we study $\Sigma$ versus $\left<m\right>_{\ell_1 \ell_2}$ for both the normal and 
inverted hierarchies. To do this in a comprehensive manner, 
we carry out a Monte Carlo sampling of the oscillation parameters and phases. 
The Dirac phase ($\delta$) and the two Majorana phases ($\phi_2$ and $\phi_3$) 
are allowed to range between $0$ and 2$\pi$.  
The atmospheric oscillation data gives bounds on  \cite{SuperK} 
$$ 1.9\times 10^{-3} \mbox { eV}^2 <| \delta m^2_a| < 3.0\times10^{-3} \mbox { eV}^2,\  
\sin^2 2\theta_a> 0.9\ {\rm at}\  90\%\ {\rm CL}.$$ 
In fact, the bounds on $\theta_a$ vary with $\delta m^2_a$ and in our computation 
these bounds are obtained from the 90$\%$ CL $\delta m^2_a$ versus 
sin$^2 2\theta_a$ plot of Ref.~\cite{Kearns}, which was obtained in an $L/E$ analysis of only selected high resolution FC (fully contained) and PC (partially contained) events. The analysis of the full data set from the same running period \cite{SKNew} gives slightly different constraints; the constraints on sin$^2 2\theta_a$ are slightly better from the full data set but the $L/E$ 
analysis better constrains $\delta m^2_a$. Similarly, for the above range of $\delta m^2_a$, reactor data places bounds on
$$\sin^2\theta_x< 0.06,$$ 
which also vary with $\delta m^2_a$. 
We use the limits obtained from the CHOOZ 90$\%$ CL exclusion plot of Ref.~\cite{CHOOZ}. 
Finally a joint analysis of the solar and reactor oscillation data limits the parameter ranges to
$$7.6\times10^{-5}\  \mbox{eV}^2 < \delta m^2_s <
9.1\times10^{-5}\  \mbox {eV}^2,\  {\rm and}\  
0.31 < \tan^2\theta_s < 0.52\ {\rm  at}\ 90\% \ {\rm CL.}$$

The bounds on $\theta_s$ vary with $\delta m^2_s$ and these are obtained from the 90$\%$ 
CL $\delta m^2_s$ versus tan$^2 \theta_s$ plot of Ref.~\cite{Bahcall}. 
All the inputs to the Monte Carlo sampling are summarized in Table I. 
We can further constrain the range of $\left<m\right>_{\ell_1 \ell_2}$ by imposing 
limits on $\Sigma$ obtained from cosmology.  The current best limit 
\begin{equation}
\Sigma \le \mbox{0.42 eV at } 95\% \mbox{ CL}             
\label{CosmLim}
\end{equation}
 was obtained from an analysis of WMAP, the SDSS galaxy spectrum and its bias, the SDSS Ly$\alpha$ forest power spectrum and the latest supernovae SNIa sample, 
 assuming a spatially flat universe and adiabatic initial conditions \cite{Seljak}.  Throughout this
 paper, we will adopt the cosmological bound of Eq.~(\ref{CosmLim}). We would like to point
 out that more conservative analyses without including the Ly$\alpha$ forest power spectrum exist. These lead to larger values of $\Sigma$, such as $\Sigma \le \mbox{0.75 eV}$ \cite{Adam} and $\Sigma \le \mbox{0.54 eV}$ \cite{OtherSeljak} at the 2$\sigma$ level. 

\begin{table}[tb]
\caption{Inputs to Monte Carlo sampling for calculating effective neutrino masses}
\label{table_MCinput}
\begin{center}
\begin{tabular}{|c|c|}
\hline
Parameter & Input \\
\hline
$|\delta m_a^2|$  &  $1.9\times10^{-3} \mbox { eV}^2 - 3.0\times10^{-3} \mbox { eV}^2$ \cite{SuperK}    \\
$\delta m_s^2$  & 90$\%$ CL $\delta m^2_s$ versus tan$^2 \theta_s$ plot \cite{Bahcall}     \\
$\theta_a$ & 90$\%$ CL $\delta m^2_a$ versus sin$^2 2\theta_a$ plot \cite{Kearns} \\
$\theta_s$ & 90$\%$ CL $\delta m^2_s$ versus tan$^2 \theta_s$ plot \cite{Bahcall}    \\
$\theta_x$ &CHOOZ 90$\%$ CL exclusion plot  \cite{CHOOZ} \\
$\delta$ & 0 to 2$\pi$     \\
$\phi_2$ & 0 to 2$\pi$     \\
$\phi_3$ & 0 to 2$\pi$     \\
\hline
\end{tabular}
\end{center}
\end{table}
 
 The results of the Monte Carlo sampling along with constraints from cosmology 
 are shown in Fig.~\ref{fig:1} for both normal and inverted hierachies by solid and
 dashed curves respectively.  We first point out  the following  general  features: 
\begin{itemize}
\item
For  large  values of the minimum neutrino mass, typically
$m_{min}>0.1$ eV, the mass differences are unimportant and thus
$m_1 \approx m_2 \approx m_3$ or $\Sigma \gsim 3m_{min}$. 
Due to unitarity of the mixing matrix, we can see 
that the maximum values of effective masses obey 
$\left<m\right>_{\ell_1 \ell_2}\approx m_{min} \lsim \Sigma/3.$  
On the other hand, for smaller values of $m_{min}$,  
$\Sigma$  is governed  by the larger mass difference, namely 
 $\Sigma \gsim \sqrt{\delta m^2_a} \approx m_3\approx 5\times 10^{-2}$ eV  for
 the NH scenario, and $\Sigma \gsim 2\sqrt{\delta m^2_a} \approx 
 2m_1\approx 9\times 10^{-2}$ eV for the IH scenario.

\item
The  cosmological observations of $\Sigma \le$ 0.42 eV 
put  an upper limit on $\left<m\right>_{\ell_1\ell_2}$ at about 0.14 eV.
\item
The normal and inverted hierarchies are indistinguishable at the current level of sensitivity.
When $\Sigma$ is smaller than the present cosmology limit,  the difference (in upper limits) between normal and inverted 
hierarchies may become significant as is apparent from all the plots. 
We are just above the region where the NH and IH results begin to separate. An improvement
in the accuracy of either of the two observables by a factor of two or better
would begin to provide a  sensitive probe to distinguish the NH and IH scenarios. 
\end{itemize}
We note the qualitative difference between Fig.~\ref{fig:1}(a) and the other panels in Fig.~\ref{fig:1}.
The allowed region for ${\left<m\right>_{ee}}$ is between the curves 
for both normal hierarchy (solid lines) and inverted hierarchy (dashed curves).  
It is more stringent than the other ${\left<m\right>_{\ell_1 \ell_2}}$ because ${\left<m\right>_{ee}}$ depends on
two fewer parameters ($\theta_a,\ \delta$) than the others. Also the specific combination of oscillation parameters for ${\left<m\right>_{ee}}$ does not lead to complete cancellations or vanishingly small contributions (unless for a small range of $\Sigma$ in the NH scenario) unlike others.
Our results for the range of values of ${\left<m\right>_{ee}}$ are similar to the analyses 
previously done in Refs.~\cite{Glashow,Pas} and 
an updated version presented in Ref.~\cite{Fogli}.
Similar analyses for the other $\left<m\right>_{\ell_1\ell_2}$ for specific scenarios 
were considered in Ref.~\cite{WR}.
We will not show the results for  $\left<m\right>_{e\mu}$ and  
$\left<m\right>_{\tau\tau}$ since they are qualitatively very similar to 
 $\left<m\right>_{e\tau}$ and  $\left<m\right>_{\mu\mu}$, respectively.
 
By setting limits on the $\lv$  process decay rates and cross sections, 
direct experimental upper bounds on effective neutrino masses can be obtained 
as to be discussed in the next sections. We then compare them with the results from this section.

%%%%%%%%%%%%%%%%%%%%%%%%%%%%%%%%%%%%%%%%%%%%%%%%%%%%%%%%%%%%%%%
\section{Neutrinoless double-beta decay $(0\nu\beta\beta)$}
\label{ee0n}
%%%%%%%%%%%%%%%%%%%%%%%%%%%%%%%%%%%%%%%%%%%%%%%%%%%%%%%%%%%%%%%

\begin{figure}
\includegraphics[width=3.15in]{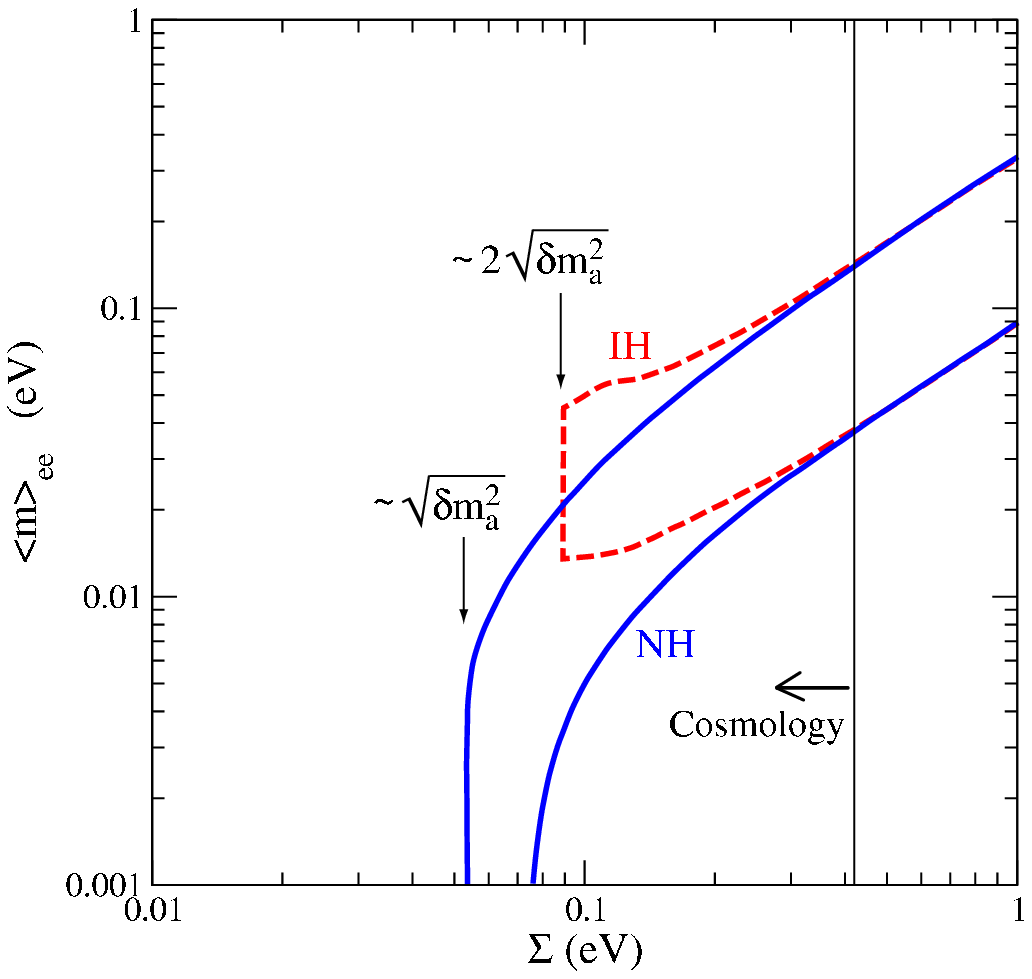}\hfill
\includegraphics[width=3.15in]{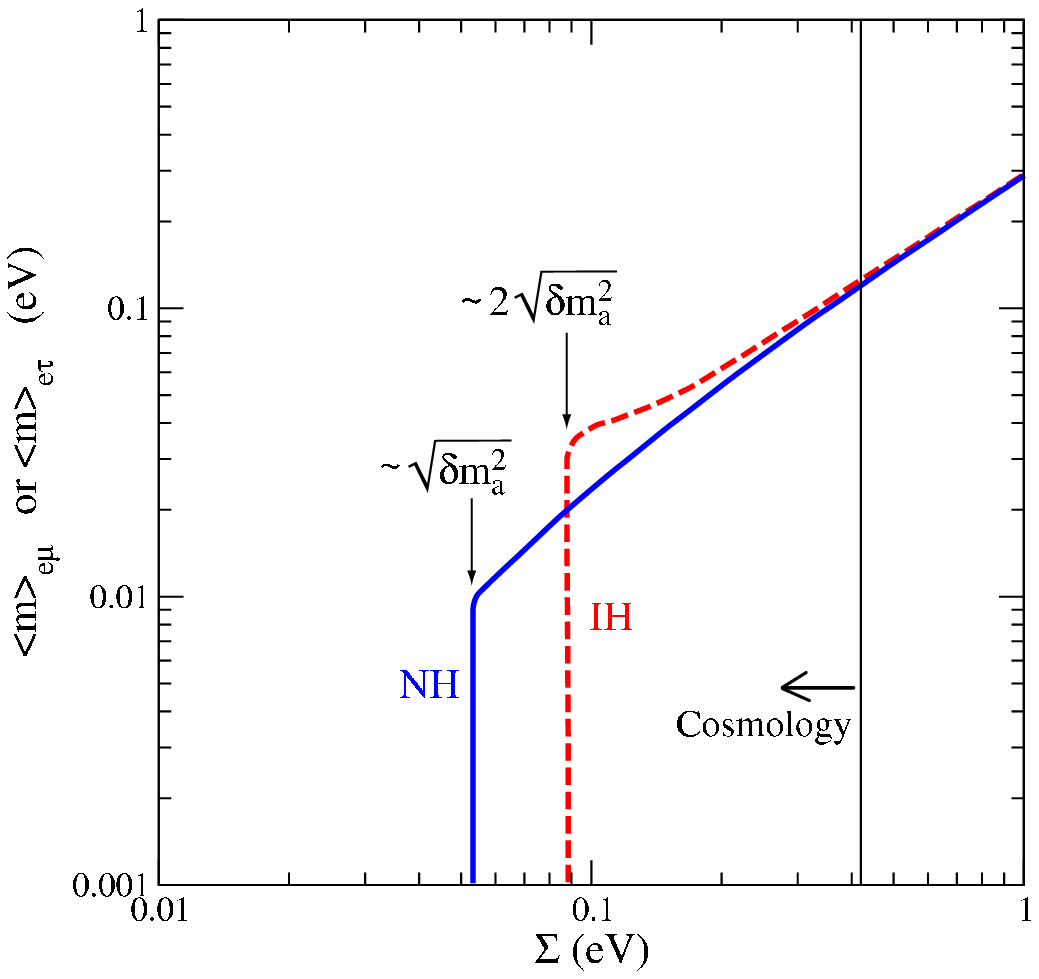} \\[.1in] 
\includegraphics[width=3.15in]{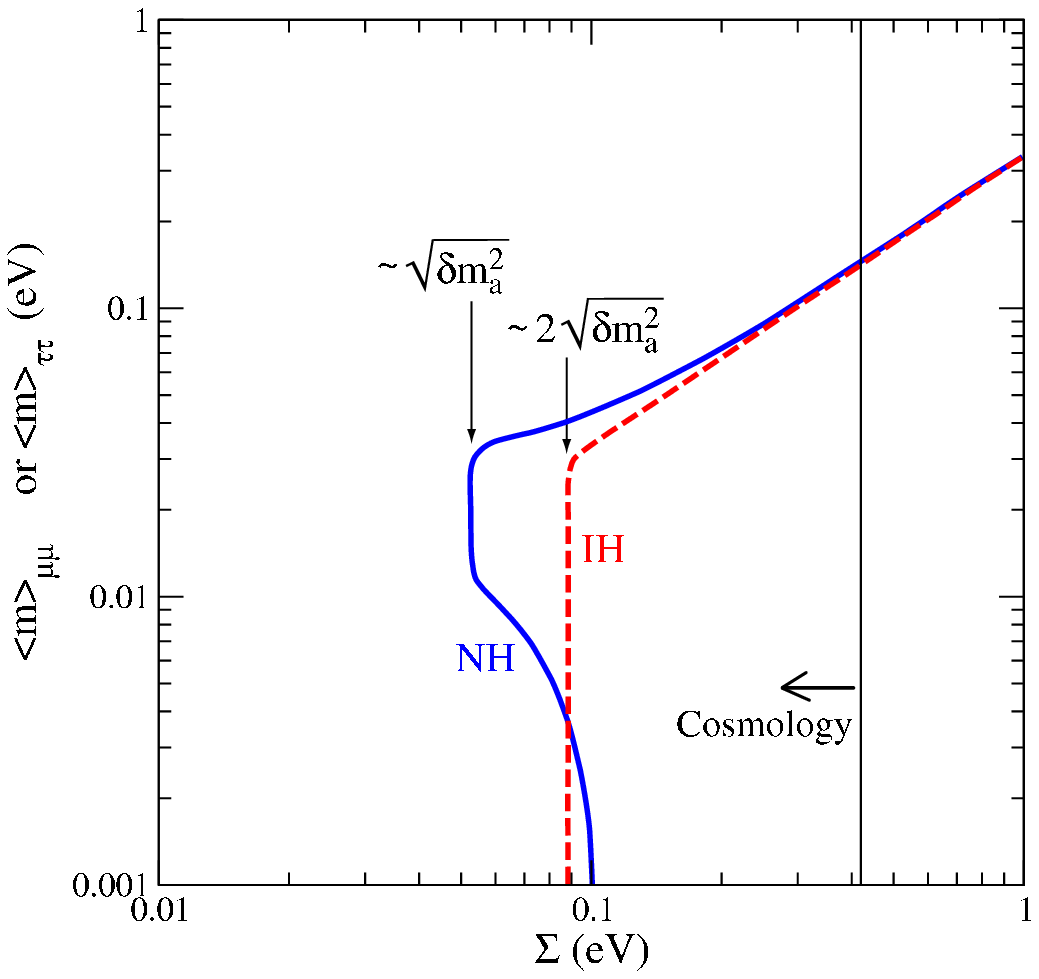}\hfill
\includegraphics[width=3.15in]{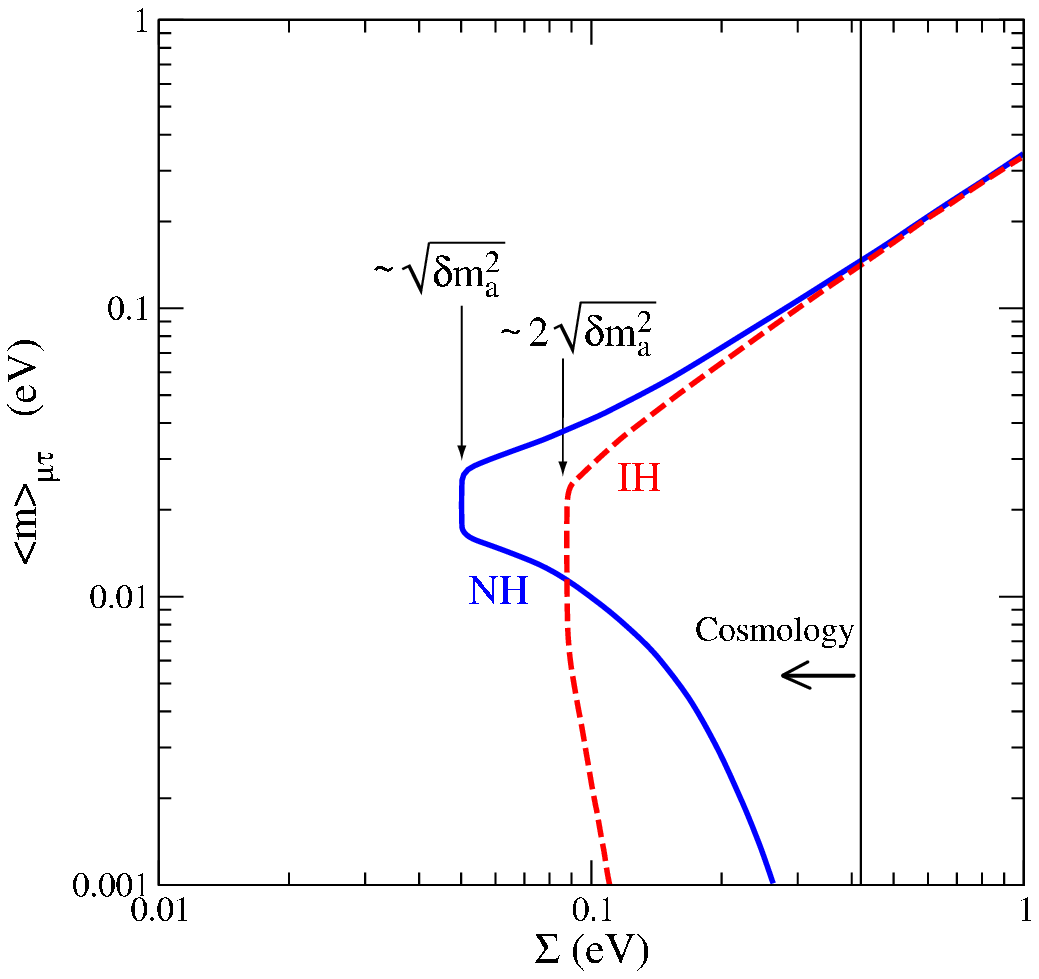}
\caption{%
(a) Upper left: allowed regions between the curves for $\left<m\right>_{ee}$ versus  $\Sigma$; 
(b) upper right: allowed regions to the right of the curves for  $\left<m\right>_{e\mu}$ or $\left<m\right>_{e\tau}$; 
(c)  lower left: same as (b) but for $\left<m\right>_{\mu\mu}$ or $\left<m\right>_{\tau\tau}$; 
(d) lower right: same as (b) but  for  $\left<m\right>_{\mu\tau}$. 
The bound for $\Sigma$ at 95$\%$ CL  from cosmology is shown in all figures by the vertical line.}
\label{fig:1} 
\end{figure}

The decay rate for neutrinoless double-beta decay $(0\nu\beta\beta)$ 
is proportional to ${\left<m\right>^2_{ee}}$; ${\left<m\right>_{ee}}$ is plotted in  Fig.~\ref{fig:1}(a).
The theoretical formalism for calculating the decay rate 
is given in Appendix \ref{appB}.
Ref.~\cite{ElliottEngel} summarized the latest experimental limits of  $0\nu\beta\beta$ 
for various isotopes; the results are reproduced in Table \ref{table_doublebetadecay}.  
The experimental bound on $\left<m\right>_{ee}$ has improved from 5 eV 
in 1992 to about 1 eV in the most recent experiments. The best limits come from the 
two $^{76}$Ge experiments which are Heidelberg-Moscow and IGEX respectively. 
Although $0\nu\beta\beta$ is a vital experiment to determine the Majorana nature of  neutrinos, 
we note that the uncertainty in nuclear matrix elements would result in an uncertainty 
as large as a factor of three in the inferred value of $\left<m\right>_{ee}$ from an observation of the decay process \cite{ElliottEngel}. 

We list our result in the last row of Table \ref{table_doublebetadecay} based on  Fig.~\ref{fig:1}(a)
which is largely determined by the cosmological bound. 
We see that our result is slightly stronger than the current experimental bounds.
The bounds we obtain for $\left<m\right>_{ee}$ are 
based on the Monte Carlo scan over the fundamental parameters in the neutrino
sector as given in Table \ref{table_MCinput} 
and thus are independent of the nuclear matrix elements.
The large uncertainty comes in when we predict the decay rates for the nuclear isotopes.

\begin{table}[tb]
\caption{Experimental bounds on half-life time of $0\nu\beta\beta$ for various isotopes
from Ref.~\cite{ElliottEngel}  and the implied upper bounds on $\left<m\right>_{ee}$.
The last row lists the cosmological bound as obtained in the previous section.}
\label{table_doublebetadecay}
\begin{center}
\begin{tabular}{|c|c|c|c|}
\hline
Isotope & Half-life (years)      & $\left<m\right>_{ee}$ (eV)      & Year of published paper \\
\hline
$^{48}$Ca & $>1.4 \times 10^{22}$   & $< 7.2 - 44.7$  & 2004 \\
\hline
$^{76}$Ge & $ > 1.9 \times 10^{25}$    & $ < 0.35$ & 2001 \\
\hline
$^{76}$Ge & $ > 1.6 \times 10^{25}$   & $ < 0.33 - 1.35$  & 2002 \\
\hline
$^{76}$Ge & $ = 1.2 \times 10^{25}$   & $ = 0.44$  & 2004 \\
\hline
$^{82}$Se & $ > 2.7 \times 10^{22}$   & $ < 5$ & 1992 \\
\hline
$^{100}$Mo & $ > 5.5 \times 10^{22}$    & $ < 2.1$  & 2001 \\
\hline
$^{116}$Cd & $ >1.7 \times 10^{23}$    & $ < 1.7$ & 2003 \\
\hline
$^{128}$Te & $ > 7.7 \times 10^{24}$  & $ < 1.1 - 1.5$  & 1993 \\
\hline
$^{130}$Te & $ > 5.5 \times 10^{23}$  & $ < 0.37 - 1.9$ & 2004 \\
\hline
$^{136}$Xe & $ > 4.4 \times 10^{23}$ & $ < 1.8 - 5.2$  & 1998 \\
\hline
$^{150}$Nd & $ > 1.2 \times 10^{21}$  & $ < 3$  & 1997 \\
\hline
Cosmology &  none  & $ \le 0.14$  & this paper \\
\hline
\end{tabular}
\end{center}
\end{table}

A recent publication claims evidence for $0\nu\beta\beta$ at the $4.2\sigma$ 
level \cite{Klapdor}, but the result is controversial.  In Ref.~\cite{Klapdor}, 
 $\left<m\right>_{ee}$ was determined to be in a range  between 
 0.2 to 0.6 eV at  99.73$\% $ CL for a particular choice of the nuclear matrix 
element, and  becomes 0.1 to 0.9 eV if allowing a  $\pm50\% $ 
uncertainty of the nuclear matrix element.
With the bound on $\Sigma$ at 95$\%$ CL, the first  range is disfavored by the limits 
we obtained and the second range allowing a larger uncertainty leaves a very 
narrow range for ${\left<m\right>_{ee}}$. 

Tritium $\beta$ decay experiments also probe the absolute scale of the neutrino mass. The current limits from cosmology are better by an order of magnitude compared to the tritium $\beta$ decay limits \cite{Fogli}. Future limits from the KATRIN tritium $\beta$-decay experiment are expected to be 
0.30 eV (0.35 eV) at the 3$\sigma$ (5$\sigma$) level.  
The present limit on $\left<m\right>_{ee}$ from cosmology 
is still stronger than this expected improvement from KATRIN, 
but the KATRIN experiment will provide an important direct confirmation.

%%%%%%%%%%%%%%%%%%%%%%%%%%%%%%%%%%%%%%%%%%%%%%%
\section{Lepton-number violating tau decay}
\label{taud}
%%%%%%%%%%%%%%%%%%%%%%%%%%%%%%%%%%%%%%%%%%%%%%%

In this section we examine tau decays into an anti-lepton and two mesons 
\begin{equation}
\tau^- \rightarrow \ell^+ M_1^- M_2^-, 
\label{taudec}
\end{equation}
which is a process with $\Delta L=-2$. 
The  relevant effective masses are  $\left<m\right>_{e\tau}$ and $\left<m\right>_{\mu\tau}$
as shown in Fig.~\ref{fig:1}(b) and Fig.~\ref{fig:1}(d) respectively, from the current
parameters from neutrino oscillation experiments. 
The constraint from cosmology of Eq.~(\ref{CosmLim}) gives an upper limit of 0.14 eV for $\left<m\right>_{e\tau}$ and $\left<m\right>_{\mu\tau}$.
In Appendix \ref{appC}, we give the calculations for the decay branching fraction
of the process (\ref{taudec}) in terms of $\left<m\right>^2_{\ell \tau}$.
We express the branching fraction in an intuitive form as
\begin{eqnarray}
BR \approx  10^{-33}\  |V^{CKM}_{M_1} V^{CKM}_{M_2}|^2  
\left({f_{M_1}\ f_{M_2}}\over{(100\ {\mev})^2}\right)^2 
\left({1777\ {\mev}}\over{m_{\tau}}\right)^2 
\left({\left<m\right>_{\ell \tau}}\over{1\ {\ev}}\right)^2 \Phi,
\label{br}
\end{eqnarray}
where $\Phi$ is the phase space integral over the squared matrix element
and can be evaluated numerically.
For small values of $\left<m\right>_{\ell \tau}$, the branching fraction induced
by a light Majorana neutrino is seen to be very small.

A direct search for neutrinoless tau decays has been made at the CLEO II detector at Cornell Electron Storage Ring (CESR). 
Twenty eight different decay modes have been studied and  
the limits on the branching fractions were reported in \cite{CLEO}. 
  The experimental limits for various decay modes are typically of the order of
  $10^{-6}$, as   given in Table \ref{table_taudecay}.
From those, one can determine upper bounds on $\left<m\right>_{\ell\tau}$
from Eq.~(\ref{br}), as given in details  in Appendix \ref{appC}.
Unfortunately, the obtained bounds are very weak  compared
  to our inferred cosmology bounds less than 1 eV, as shown in the last
  column in Table \ref{table_taudecay}.
  In fact,   the current formalism for calculations in terms of the 
  effective neutrino masses 
  is valid only  for the light Majorana neutrino exchange  
  when the mass is much less than the energies available in the reaction. 
  Thus the entries with such large values in this Table lose the
  original meaning of the effective neutrino mass.
  We nevertheless include these values here and henceforth  to indicate how much
  improvement would be needed to be sensitive to the light neutrino contributions. 
  This information is useful to see which process may be more sensitive
  to what operator and to what extent.   On the other hand, any 
   observation of a $\lv$ signal in these channels at the current values would strongly imply
    contributions beyond those of light Majorana neutrinos. Hence it is important not to neglect the experimental study of these processes even though the limits seem too weak in comparison with the Majorana neutrino mechanism. 
 
\begin{table}[tb]
\caption{Experimental bounds on branching fractions in $\dl=2$
tau decays from \cite{CLEO}
and the implied sensitivity to probe the corresponding  effective neutrino masses.}
\label{table_taudecay}
\begin{center}
\begin{tabular}{|c|c|c|}
\hline
Decay mode & $B_{exp}$      & $\left<m\right>_{\ell\tau}$ (TeV) \\
\hline
$\tau^- \rightarrow e^+ \pi^- \pi^-$  & $1.9 \times 10^{-6}$   & $12$ \\
\hline
$\tau^- \rightarrow e^+ \pi^- K^-$  & $2.1 \times 10^{-6}$    & $46$  \\
\hline
$\tau^- \rightarrow e^+ K^- K^-$ & $3.8 \times 10^{-6}$   & $730$  \\
\hline
$\tau^- \rightarrow \mu^+ \pi^- \pi^-$ & $ 3.4 \times 10^{-6}$   & $20$ \\
\hline
$\tau^- \rightarrow \mu^+ \pi^- K^-$ & $7.0 \times 10^{-6}$    & $100$ \\
\hline
$\tau^- \rightarrow \mu^+ K^- K^-$ & $6.0 \times 10^{-6}$    & $1000$ \\
\hline
\end{tabular}
\end{center}
\end{table}   

\section{Rare meson decays}
\label{mesond}

We now investigate the $\lv$ processes in which a meson decays  \cite{Littenberg, AliBorisov}
into another meson and two like-sign leptons 
\begin{equation}
M_1^+ \rightarrow M_2^- \ell_1^+ \ell_2^+.
\end{equation}
These decays are  similar to the tau decay modes described in the last section. 
For the various decay modes, the effective neutrino masses involved are
 $\left<m\right>_{ee}$, $\left<m\right>_{e\mu}$ and $\left<m\right>_{\mu\mu}$
 depending on the final state leptons. Again, we plot their allowed values based on the
known neutrino parameters,   as shown in Fig.~\ref{fig:1}(a) for  $\left<m\right>_{ee}$,
 in Fig.~\ref{fig:1}(b) for  $\left<m\right>_{e \mu}$
 (indistinguishable from  $\left<m\right>_{e \tau}$),
and  in Fig.~\ref{fig:1}(c) for $\left<m\right>_{\mu\mu}$
 (indistinguishable from  $\left<m\right>_{\tau \tau}$).
We infer a generic  upper limit for  $\left<m\right>_{\ell_1\ell_2}$ to be 0.14 eV from  constraints from cosmology Eq.~(\ref{CosmLim}). The branching fraction for the rare meson decay modes is 

\begin{eqnarray}
BR  \approx 10^{-29}\ |V^{CKM}_{M_1} V^{CKM}_{M_2}|^2 \left({\tau_{M_1}}
\over{1.0\times 10^{-8}\ {s}}\right)
\left({f_{M_1}\ f_{M_2}} \over{(100\ {\mev})^2}\right)^2 
 \left({m_{M_1}}\over{1\ {\gev}}\right)^3 \left({\left<m\right>_{\ell_1 \ell_2}}\over{1\ {\ev}}\right)^2 \Phi', ~~~
\end{eqnarray}
where $\Phi'$ is the phase space integral over the squared matrix element
and can be evaluated numerically.

Searches for rare meson decay modes have been made in numerous experiments. Table \ref{table_raremesondecay} summarizes the current experimental limits on branching 
fractions given by \cite{PDG}. From these, direct search 
limits can be determined on effective neutrino masses and
some associated calculational  details are given in Appendix \ref{appD}. 
Again,  the bounds obtained are still much weaker than the 
 cosmology bound. 
Although the $K^+$ decays yield the most sensitive bounds, they are still
many orders of magnitude away.   We include the obtained values 
  in Table \ref{table_raremesondecay} to indicate how much
  improvement would be needed to be sensitive to the light neutrino contributions. There are no direct search 
limits obtained for ${\langle m \rangle}_{\tau \tau}$ from the processes discussed.
However only very weak constraints for 
$BR(B \to X \tau^+ \tau^-) < \cal{O}(\mbox{5}\%)$ exist in a theoretical analysis \cite{Grossman}.  
The similar
signature $B^+ \rightarrow M^- \tau^+ \tau^+$ is a  possible decay mode that would bound 
  ${\langle m \rangle}_{\tau \tau}$ and should be pursued,
  but any such bound will not be competitive with the cosmology limit unless
  there is a contribution from new physics beyond the light Majorana neutrinos.

\begin{table}[tb]
\caption{Experimental bounds on branching fractions in  $\dl=2$
rare meson decays \cite{PDG}
and the implied  sensitivity to probe the corresponding effective neutrino masses.}
\label{table_raremesondecay}
\begin{center}
\begin{tabular}{|l|l|c|}
\hline
Decay mode & $B_{exp}$      & $\left<m\right>_{\ell_1 \ell_2}$ (TeV)  \\
\hline
$K^+ \rightarrow \pi^- e^+ e^+$ & $6.4 \times 10^{-10}$   & 0.11  \\
$K^+ \rightarrow \pi^- \mu^+ \mu^+$ & $3.0 \times 10^{-9}$   &  0.48 \\
$K^+ \rightarrow \pi^- e^+ \mu^+$ & $5.0 \times 10^{-10}$   & 0.09  \\
\hline
$D^+ \rightarrow \pi^- e^+ e^+$ & $9.6 \times 10^{-5}$   & 320  \\
$D^+ \rightarrow \pi^- \mu^+ \mu^+$ & $4.8 \times 10^{-6}$   & 76   \\
$D^+ \rightarrow \pi^- e^+ \mu^+$ & $5.0 \times 10^{-5}$   & 170  \\
\hline
$D^+ \rightarrow K^- e^+ e^+$ & $1.2 \times 10^{-4}$   & 1900  \\
$D^+ \rightarrow K^- \mu^+ \mu^+$ & $1.3 \times 10^{-5}$   & 670  \\
$D^+ \rightarrow K^- e^+ \mu^+$ & $1.3 \times 10^{-4}$   & 1500  \\
\hline
$D_s^+ \rightarrow \pi^- e^+ e^+$ & $6.9 \times 10^{-4}$   & 200  \\
$D_s^+ \rightarrow \pi^- \mu^+ \mu^+$ & $2.9 \times 10^{-5}$   &42   \\
$D_s^+ \rightarrow \pi^- e^+ \mu^+$ & $7.3 \times 10^{-4}$   & 150  \\
\hline
$D_s^+ \rightarrow K^- e^+ e^+$ & $6.3 \times 10^{-4}$   &990   \\
$D_s^+ \rightarrow K^- \mu^+ \mu^+$ & $1.3 \times 10^{-5}$   & 150  \\
$D_s^+ \rightarrow K^- e^+ \mu^+$ & $6.8 \times 10^{-4}$   & 740 \\
\hline
$B^+ \rightarrow \pi^- e^+ e^+$ & $1.6 \times 10^{-6}$   & 420  \\
$B^+ \rightarrow \pi^- \mu^+ \mu^+$ & $1.4 \times 10^{-6}$   & 400  \\
$B^+ \rightarrow \pi^- e^+ \mu^+$ & $1.3 \times 10^{-6}$   & 270  \\
\hline
$B^+ \rightarrow K^- e^+ e^+$ & $1.0 \times 10^{-6}$   & 1300  \\
$B^+ \rightarrow K^- \mu^+ \mu^+$ & $1.8 \times 10^{-6}$   & 1800  \\
$B^+ \rightarrow K^- e^+ \mu^+$ & $2.0 \times 10^{-6}$   & 1300 \\
\hline
\end{tabular}
\end{center}
\end{table}

%%%%%%%%%%%%%%%%%%%%%%%%%%%%%%%%%%%%%%%%%%%%%%%% 
\section{Muon positron conversion }
\label{mue}
%%%%%%%%%%%%%%%%%%%%%%%%%%%%%%%%%%%%%%%%%%%%%%%%%%%%%

The nuclear muon to positron 
conversion process is another $\dl=2$ process that is very similar to $0\nu\beta\beta$. When a muon propagates through matter, ordinarily it interacts with a proton in a nucleus and produces a 
neutron and a neutrino, which is similar to inverse beta decay. However, if the neutrino is a Majorana particle, it is 
possible that a muon can interact with two protons and produce two neutrons and a positron. The leptonic part of the decay amplitude is exactly the same as tau decay and the nuclear part will lead to nuclear matrix elements analogous to $0\nu\beta\beta$. The fundamental interaction is parameterized
by $\left<m\right>_{e \mu}$ and the current bound from neutrino oscillation experiments and
cosmology is shown  in Fig.~\ref{fig:1}(b).

An experimental bound on the branching ratio of muon to positron conversion 
on titanium was reported in  \cite{SINDRUM}
\begin{equation}
\label{31}
B = \frac {\Gamma (Ti + \mu^- \rightarrow e^+ + Ca_{gs})} {\Gamma (Ti + \mu^- \rightarrow \nu_\mu + Sc)} <  1.7 \times 10^{-12}.
\end{equation}
The experimental limit on $\left<m\right>_{e \mu}$ was obtained from this branching ratio limit 
by Ref.~\cite{Zuber} to be 
\begin{equation}
\nonumber
\left<m\right>_{e \mu} \le 17\ (82)\  \mev,
\end{equation}
 where the created proton pairs are in singlet (triplet) state. Although still larger than the
 bound from oscillation plus cosmology, this can be the next most sensitive probe to the $\lv$
 processes after $0\nu\beta\beta$.
 However, others \cite{Simkovic} argue that the theoretical expression for the decay rate was overestimated and they obtain a much lower branching ratio prediction,
\begin{equation}
\label{32}
B = 1.60 \times 10^{-25}{\biggl (\frac {\left<m\right>_{e \mu}}{m_e}\biggr )}^2,
\end{equation}
which would only lead to a weak bound of $\left<m\right>_{e \mu} \le 1.3$ TeV.
Thus there is a large disparity of $\sim 10^5$ 
in the literature about the inferred limits on 
$\left<m\right>_{e \mu}$ from the muon-positron conversion process,
due to the different  treatments of the nuclear transition matrix elements. 
Beside this difference, there is a large uncertainty in muon-positron conversion 
due to the effects from nuclear physics. 
As a competing channel, the limit from $K^+$ decay of 90 GeV is more constraining 
than this 1.3 TeV limit but weaker than the optimistic result above. Moreover, the hadronic matrix
element for the kaon decay should be better known than that involving nuclear physics.

Another process similar to muon-positron conversion that has not been studied experimentally so far is nuclear muon capture: $\mu^- + (Z,A) \to \mu^+ + (Z-2,A)$. This process was first studied theoretically for $^{44}Ti$ and a branching ratio of 
$5.0 \times 10^{-24}$ was obtained by considering an effective neutrino mass 
$\left<m\right>_{\mu \mu}$ of 250 keV \cite{Missimer}. By including our limit from cosmology of $\left<m\right>_{\mu \mu} \le$ 0.14 eV, we can deduce a branching ratio $\le 1.6 \times 10^{-36}$. 
Ref.~\cite{Faessler} claims that the imaginary part of the nuclear matrix elements 
which plays a dominant role was neglected in \cite{Missimer} which led to an overestimation. 
They obtain a branching ratio
\begin{equation}
\label{33}
B = 1.0 \times 10^{-23}{\biggl (\frac {\left<m\right>_{\mu \mu}}{m_e}\biggr )}^2.
\end{equation}
With our limit of $\left<m\right>_{\mu \mu} \le$ 0.14 eV this translates to a branching ratio 
$\le 0.75 \times 10^{-36}$. The most recent paper on this topic \cite{Takasugi} claims the branching 
ratio is $10^{-6}$ smaller than the one estimated in \cite{Missimer} and would lead to a branching ratio $\le 10^{-42}$ if we consider  $\left<m\right>_{\mu \mu} \le$ 0.14 eV. Evidently there is a disagreement in the literature about the limit on the branching ratio by 6 orders of magnitude. Only an effort similar to that for $0\nu\nu\beta$ can improve the situation. The $\mu^- \to \mu^+$ conversion process will be studied experimentally at the PRISM facility and is expected to achieve sensitivity of $\sim 10^{-13}$ to this process on $^{44}Ti$ nucleus after one nominal year run \cite{Aoki}. But this is much below the predicted branching ratios even for the most optimistic scenario and will not be accessible in the near future if the process is mediated by light Majorana neutrinos only.

\section{Conclusion}
\label{concld}

\begin{table}[tb]
\caption{Summary of experimental bounds and cosmology limits on 
effective neutrino mass. The lowest bounds on each 
component of effective neutrino mass are presented with the corresponding 
experiments for these bounds.}
\label{table_summary}
\begin{center}
\begin{tabular}{|c|c|c|l|}
\hline
$\ell_1 \ell_2$ & Cosmo bounds on $\left<m\right>_{\ell_1 \ell_2}$  & Exp bounds on $\left<m\right>_{\ell_1 \ell_2}$   & Corresponding experiments \\
\hline
$ee$ & 0.14 eV & 0.33 eV & $0\nu\beta\beta$ \\
\hline
$e\mu$ & 0.14 eV & 17 MeV (90\ GeV)$^\dagger$ & $\mu^- - e^+$ conversion \\ 
\hline
$e\tau$ & 0.14 eV  & 12 TeV & $\tau^- \rightarrow e^+ \pi^- \pi^-$ \\
\hline
$\mu\mu$ & 0.14 eV & 480 GeV  & $K^+ \rightarrow \pi^- \mu^+ \mu^+$ \\
\hline
$\mu\tau$ & 0.14 eV  & 19 TeV  & $\tau^- \rightarrow \mu^+ \pi^- \pi^-$ \\
\hline
$\tau\tau$ & 0.14 eV & none  & none \\
\hline
\multicolumn{4}{l}{$^\dagger$ The conservative limit comes from $K^+ \rightarrow \pi^- e^+ \mu^+$ 
which, unlike $\mu^--e^+$ conversion,}\\
\multicolumn{4}{l}{ does not involve the large uncertainties from nuclear matrix element calculations.}
\end{tabular}
\end{center}
\end{table}

The observation of a $\lv$ process would show that neutrino is a Majorana particle. 
In the absence of exotic $\lv$ interactions, the rates for these processes 
are determined by effective neutrino masses $\left<m\right>_{\ell_1\ell_2}$, 
as functions of light Majorana neutrino masses and the mixing parameters.
We first sampled the experimentally allowed ranges of $\left<m\right>_{\ell_1\ell_2}$ 
based on the data from neutrino oscillation experiments as well as 
cosmological observations,  and obtained a stringent upper bound 
$\left<m\right>_{\ell_1\ell_2} \approx \Sigma/3 \lsim 0.14$ eV. 
This cosmology limit is expected to improve with a future sensitivity down to 
$\Sigma \approx 0.1$ eV \cite{WayneHu}. As the limits on $\Sigma$ improve 
new bounds on ${\langle m \rangle}_{\ell_1 \ell_2}$ can be deduced as seen from our plots.
In particular, the normal hierarchy and inverted hierarchy scenarios may be experimentally differentiated.

We  considered  four lepton-number violating processes: 
(a) neutrinoless double-beta decay $(0\nu\beta\beta)$, (b) $\dl=2$ tau decays, 
(c) $\dl=2$ rare meson decays and (d) nuclear muon-positron conversion. 
After evaluating the transition rates for these processes, we translated  the
current experimental  bounds from direct searches into limits on 
$\left<m\right>_{\ell_1\ell_2}$.
The best limits  obtained  from experiments on 
these processes were compared with the cosmology limits 
in Table \ref{table_summary}.  The $0\nu\beta\beta$ process is the only process which can currently
  give interesting experimental limits on ${\langle m \rangle}_{\ell_1 \ell_2}$. We note that while experimental limits for  $0\nu\beta\beta$ involve large theoretical uncertainties from nuclear matrix element calculations, our cosmology limit   is independent of any such uncertainties. 
  The other processes have very weak experimental limits, that essentially  do not impose
  any meaningful bounds on   ${\langle m \rangle}_{\ell_1 \ell_2}$.
  The entries in the Tables are only meant to suggest the level of improvement
needed in order to sensitively probe the light Majorana neutrino mass. 
On the other hand, the predicted small rates could provide a window of
opportunity for observing exciting new physics.
Any positive signal in those direct searches would indicate new contributions 
to the $\lv$ interactions beyond those from the three light Majorana neutrinos.

%%%%%%%%%%%%%%%%%%%%%%%%%%%%%%%%%%%%%%%%%%%%%%%

\vspace{1cm}

\begin{acknowledgments}
We thank Kenny Cheng for his significant participation in the early stages of this study,
and Andre de Gouvea, Yuval Grossman, Boris Kayser, and Manny Paschos for discussions. 
T.H.~would like to acknowledge the support  by a Fermilab Frontier Fellowship,
and an Argonne Scholarship. 
This research was supported in part by the U.S.~DOE under Grants 
No.DE-FG02-95ER40896, W-31-109-Eng-38,
and in part by the Wisconsin Alumni Research Foundation. 
 Fermilab is operated by the Universities Research Association 
 Inc.~under Contract No.~DE-AC02-76CH03000 with the U.S.~DOE.
\end{acknowledgments}

\vspace{1cm}

\appendix

\section{ General amplitude of $\dl=2$ processes}
\label{appA}
%%%%%%%%%%%%%%%%%%%%%%%%%%%%%%%%%%%%%%%%%%%

The charged current interaction lagrangian in terms of neutrino mass states is
\begin{equation}
\label{3}
{\cal L}_W = -\frac{g}{\sqrt{2}} \sum_{\ell = e,\mu,\tau} {\sum_{i=1,2,3}}V_{\ell i} {\bar \ell} \gamma^\mu P_L \nu_i W_\mu + h.c.
\end{equation}
where $P_L = \frac{1}{2} (1 - \gamma_5)$. The leptonic $\dl=2$ 
subprocess $W^-W^- \rightarrow \ell_1^- \ell_2^-$ is induced by
the product of  two charged  currents
\begin{equation}
\label{4}
{{\cal M}_{lep}^{\mu\nu}} \propto {\sum_i}V_{\ell_1i}V_{\ell_2i}({\bar \ell_1}
 \gamma^\mu P_L \nu_i)({\bar \ell_2} \gamma^\nu P_L \nu_i),
\end{equation}
which can be rewritten using charge conjugation as
\begin{equation}
\label{5}
{{\cal M}_{lep}^{\mu\nu}} \propto {\sum_i}V_{\ell_1 i}V_{\ell_2 i}({\bar \ell_1}
 \gamma^\mu P_L \nu_i)({\bar \nu_i}\gamma^\nu P_R { \ell^c_2}   ).
\end{equation}
The Majorana neutrino fields can be contracted to form a neutrino propagator,
and the transition matrix element is thus given by
\begin{equation}
\label{6}
{{\cal M}_{lep}^{\mu\nu}} = \frac{g^2}{2}{\sum_i}V_{\ell_1 i}V_{\ell_2 i}({\bar \ell_1}
 \gamma^\mu P_L )\frac{\qslash + m_i}{q^2 - m_i^2}(\gamma^\nu P_R { \ell^c_2}   ),
\end{equation}
where $q$ is the momentum exchange carried by the neutrino. The $\qslash$ 
term vanishes due to the chirality flip.  Including the crossed diagram ($\ell_1 \leftrightarrow \ell_2$) the leptonic amplitude then becomes
\begin{equation}
\label{9}
{{\cal M}_{lep}^{\mu\nu}} = \frac{g^2}{2}{\sum_i}V_{\ell_1 i}V_{\ell_2 i}\frac{m_i}{q^2 - m_i^2}{ \bar u_1} (\gamma^\mu \gamma^\nu + \gamma^\nu \gamma^\mu) P_R v_2  . 
\end{equation}
If we only consider the contributions from light Majorana neutrinos, 
namely $q^2\gg m_i^2$, then
\begin{eqnarray}
\nonumber
{{\cal M}_{lep}^{\mu\nu}} &=& \frac{g^2}{2}{\sum_i}V_{\ell_1 i}V_{\ell_2 i}\frac{m_i}{q^2 }{ \bar u_1} (\gamma^\mu \gamma^\nu + \gamma^\nu \gamma^\mu)P_R v_2 \\
 & = & \frac{g^2}{2} {1\over q^2}
{ \bar u_1} (\gamma^\mu \gamma^\nu + \gamma^\nu \gamma^\mu) P_R v_2\  {\sum_i}V_{\ell_1 i}V_{\ell_2 i} m_i,
\label{10}
\end{eqnarray}
and is thus governed by the ``effective neutrino mass"
\begin{eqnarray}
\nonumber
\left<m\right>_{\ell_1\ell_2} = | \sum_i V_{\ell_1 i}V_{\ell_2 i} m_i | .
\end{eqnarray}

%%%%%%%%%%%%%%%%%%%%%%%%%%%%%%%%%%%%%%%
\section{ Neutrinoless double-beta decay ($0\nu\beta\beta$)}
\label{appB}
%%%%%%%%%%%%%%%%%%%%%%%%%%%%%%%%%%%%%%%

The decay amplitude for neutrinoless double-beta decay $(0\nu\beta\beta)$ can be separated into leptonic and nuclear parts,
\begin{equation}
\label{13}
{i\cal M} = {({\cal M}_{lep})_{\mu\nu}}{({\cal M}_{nuc})^{\mu\nu}}.
\end{equation}
The leptonic amplitude is given by (\ref{10}). In the non-relativistic approximation for the nucleons, the nuclear amplitude evaluated for initial ground state to final ground state transitions turns into a sum of Gamow-Teller and Fermi nuclear matrix elements defined as,
\begin{equation}
\label{16}
{\cal M}_{nuc} \equiv {\cal M}_{GT}- \frac{g_v^2}{g_a^2}{\cal M}_{F} = \langle f| \sum_{j,k} H(r_{jk},\bar E)  \tau_j^{\dagger} \tau_k^{\dagger} 
(\stackrel{\rightarrow}{\sigma_j}\cdot \stackrel{\rightarrow}{\sigma_k}
 - \frac{g_v^2}{g_a^2}) |i \rangle ,
\end{equation}
where $\langle f|$ and $|i \rangle $ are the final and initial nuclear states, $g_a$ and $g_v$ 
are weak axial-vector and vector coupling constants and the function $H$ 
called the ``neutrino potential'' has an approximate form given in Ref \cite{ElliottEngel}. 
The decay rate for $0\nu\beta\beta$  can be expressed as
\begin{equation}
\label{15}
[T_{\frac{1}{2}}]^{-1} = G(\Delta E, Z)|{\cal M}_{nucl}|^2 \left<m\right>_{ee}^2,
\end{equation}
where $G(\Delta E, Z)$ is the phase space integral.
 
For a detailed discussion, in particular the uncertainties associated with the
nuclear matrix elements,  see \cite{ElliottEngel, DoiKotani}.

%%%%%%%%%%%%%%%%%%%%%%%%%%%%%%%%%%%%%%%%%%%%
\section{ Lepton-number violating tau decay }
\label{appC}
%%%%%%%%%%%%%%%%%%%%%%%%%%%%%%%%%%%%%%%%%%%%

This mode is cleaner in principle than $0\nu\beta\beta$ since the hadronic part does not involve complicated nuclear structure. For the tree level amplitude, the hadronic part can be expressed in terms of the decay constants of the mesons in a model independent way. The box diagram includes hadronic matrix elements which cannot be simplified in terms of decay constants and needs to be evaluated in a model dependent way. We expect the tree level amplitude to dominate and do not include the box diagram. It has been argued that in certain cases for rare meson decays sub-leading contributions may be appreciable \cite{Ivanov, AliBorisov}. Even in such a scenario the difference will not be important at the current level of sensitivities and we include the more conservative limit from tree level diagrams only. The tau decays and the rare meson decays are crossed versions of each other, hence the above arguments are true for both. 

The leptonic part of the subprocess $\tau^- \rightarrow \ell^+ W^{-*} W^{-*}$ is obtained by crossing the amplitude of $W^+ W^+ \rightarrow \ell_1^+ \ell_2^+ $ in (\ref{10})

\begin{equation}
\label{20}
{{\cal M}_{lep}^{\mu\nu}} = \frac{g^2}{2}{\sum_i}V^*_{\tau i}V^*_{\ell i}{ \bar v_\tau}\frac{m_i}{q^2} \gamma^\mu 
\gamma^\nu P_R v_\ell   .
\end{equation}
Combining the hadronic and leptonic parts, the decay amplitude for 
$$\tau^-(p_1) \rightarrow \ell^+(p_2)\  M_1^-(q_1)\  M_2^-(q_2)$$ is given by
\begin{eqnarray}
\nonumber
{i\cal M}& = &{({\cal M}_{lep})_{\mu\nu}}{{\cal M}_{M_1}^{\mu}}{{\cal M}_{M_2}^{\nu}} 
+ (M_1 \leftrightarrow M_2)\\
\label{23}
 & = & 2G_F^2 V^{CKM}_{M_1} V^{CKM}_{M_2} f_{M_1} f_{M_2} \Bigl [\sum_i
{V_{\tau i}^*}{V^*_{\ell i}}  m_i \bar v_\tau (\frac {\qslash_1 \qslash_2}{(p_1 - q_1)^2}
 + \frac {\not q_2 \not q_1}{(p_1 - q_2)^2})P_R v_\ell \Bigr ],
\end{eqnarray}
where $V^{CKM}$ is the quark flavor-mixing matrix elements for the mesons, 
$f_{M_i}$ are meson decay constants.  The decay rate is then given by
\begin{equation}
\label{LVPtauRate}
\Gamma =  (1 - {1\over 2} \delta_{M_1M_2}) \frac{1}{128\pi^3}
G_F^4 |V^{CKM}_{M_1} V^{CKM}_{M_2}|^2  f_{M_1}^2 f_{M_2}^2 {m_\tau^3}
\left<m\right>_{\ell \tau}^2 \Phi ,
\end{equation}
where $\Phi$ is the  phase space integration over the matrix elements squared
\begin{eqnarray}
\label{phi}
 \Phi &=& {1\over m_\tau^2} \int F(p_i,q_j)\ dx_\ell dx^{}_{M_1}, \\
   F(p_i,q_j) &=& {A\over (p_1-q_1)^4} + {B\over (p_1-q_1)^2 (p_1-q_2)^2} 
   + (q_1 \leftrightarrow q_2),\\
   \nonumber
      A(p_i,q_j) &=& 8(p_1\cdot q_1) (p_2\cdot q_2) (q_1\cdot q_2) -
      4m^2_{M_1} (p_1\cdot q_2) (p_2\cdot q_2)\\
\label{eq:A}
      & - & 4m^2_{M_2} (p_1\cdot q_1) (p_2\cdot q_1) + 2m^2_{M_1} m^2_{M_2} (p_1\cdot p_2),\\  
            B(p_i,q_j) &=& 4(p_1\cdot p_2) (q_1\cdot q_2)^2 -  A(p_i,q_j).
\label{eq:B}
\end{eqnarray}
The integration variables $x_\ell$ and $x_{M_1}^{}$ are the energies scaled by the mass 
of the decay particle 
\begin{equation}
\label{eq:x}
x_i={2E_i\over m_\tau},
\end{equation}
as introduced in the text book \cite{collider}. 
 Numerically, $\Phi \approx 1.641,\ 0.7787$ and 0.1455 for the  modes
$\pi \pi, \pi K$ and $KK$ respectively, neglecting the mass of the final state lepton. 

In the limit that the final state particles are massless,
then the phase space $\Phi$ can be written in a simple form as
\begin{equation}
\label{TauPhiMssless}
 \Phi = {1\over m_\tau^2} 
 \int [ 4(x_\ell - 1)+\frac{x_\ell(x_\ell - 1)}{(x_{M_1}-1)(x_\ell+x_{M_1}-1)} ] \ 
  dx_\ell dx_{M_1},
\end{equation} 
where the integration limits are given by $0 \le x_\ell \le 1$ and $(1-x_\ell) \le x_{M_1} \le 1$. Note that $\Phi$ presents a mass singularity when all the final state particles are considered massless.

Normalized to the $\tau$ decay width 
$\Gamma_\tau = G_F^2 m_\tau^5/192\pi^3$, the corresponding branching fraction is:
\begin{eqnarray}
\label{24}
BR &=&  (1 - {1\over 2} \delta_{M_1M_2})
\frac{3}{2}G_F^2 |V^{CKM}_{M_1} V^{CKM}_{M_2}|^2  f_{M_1}^2 f_{M_2}^2 \frac {1}{{m_\tau}^2} 
\left<m\right>_{\ell \tau}^2 \Phi \\
%\label{tauScldEqnA}
%&=& 6.46\times 10^{-33} |V^{CKM}_{M_1} V^{CKM}_{M_2}|^2  \left({f_{M_1}\ f_{M_2}}\over{(100\ %{\mev})^2}\right)^2 \left({1777\ {\mev}}\over{m_{\tau}}\right)^2 \left({\left<m\right>_{\ell \tau}}\over{1\ %{\ev}}\right)^2 \Phi .
&\approx & 10^{-33}\  |V^{CKM}_{M_1} V^{CKM}_{M_2}|^2  \left({f_{M_1}\ f_{M_2}}\over{(100\ {\mev})^2}\right)^2 \left({1777\ {\mev}}\over{m_{\tau}}\right)^2 \left({\left<m\right>_{\ell \tau}}\over{1\ {\ev}}\right)^2 \Phi \nonumber\\
\nonumber
&\approx&10^{-14}\ |V^{CKM}_{M_1} V^{CKM}_{M_2}|^2  \left({f_{M_1}\ f_{M_2}}\over{(100\ {\mev})^2}\right)^2  \left({\left<m\right>_{\ell \tau}}\over{m_{\tau}}\right)^2  \Phi .
\end{eqnarray}
The meson decay constants, CKM matrix elements and $\tau$ mass   
are taken from the Particle Data Group (PDG) \cite{PDG}:
$$f_\pi = 130.7\ {\mev},\ f_K = 159.8\ {\mev},\ |V_{ud}| = 0.9738,\ |V_{us}| = 0.2200.$$

%%%%%%%%%%%%%%%%%%%%%%%%%%%%%%%%%%%%%%%%%%%%%
\section{ Rare meson decay}
\label{appD}
%%%%%%%%%%%%%%%%%%%%%%%%%%%%%%%%%%%%%%%%%%%%%

The rare meson decays 
$$ M_1^+(q_1) \rightarrow \ell^+(p_1)\  \ell^+(p_2)\  M_2^-(q_2)$$ 
have the same Feynman diagrams as tau decay.
The decay amplitude is given by
\begin{equation}
\label{23a}
i{\cal M}  =  2G_F^2 V^{CKM}_{M_1} V^{CKM}_{M_2} f_{M_1} f_{M_2} \Bigl [\sum_i
{V_{\ell_1 i}}{V_{\ell_2 i}}  m_i \bar u_{\ell_1} (\frac {\qslash_1 \qslash_2}{(q_1 - p_1)^2}
 + \frac {\not q_2 \not q_1}{(q_1 - p_2)^2})P_R v_{\ell_2} \Bigr ].
\end{equation}
The decay rate is then given by:
\begin{equation}
\label{27}
\Gamma = (1 - {1\over 2} \delta_{\ell_1\ell_2})
\frac {1}{64 \pi^3} G_F^4 |V^{CKM}_{M_1} V^{CKM}_{M_2}|^2  f_{M_1}^2 f_{M_2}^2 m_{M_1}^3 \left<m\right>_{\ell_1 \ell_2}^2 \Phi'  ,
\end{equation}
where $\Phi' $ is the phase space integration with the dimensionless integration variables $x_{\ell_1}$ and $x_{M_2}$.
\begin{eqnarray}
\label{phi'}
 \Phi' &=&   \frac{1}{m_{M_1}^2}\int F'(p_i,q_j)\ dx_{\ell_1} dx_{M_2}, \\
   F'(p_i,q_j) &=& {A\over (q_1-p_1)^4} + {B\over (q_1-p_1)^2 (q_1-p_2)^2} 
   + (p_1 \leftrightarrow p_2),
\end{eqnarray}
where A$(p_i,q_j)$ and B$(p_i,q_j)$ are given in Eqs.~(\ref{eq:A}) and (\ref{eq:B}).  $x_{\ell_1}$ and $x_{M_2}$ are the energies scaled by the mass of the decay particle and are given by $x_i = 2E_i/m_{M_1}$ \cite{collider}. To have a numerical estimate consider the case when the final state particles are massless. Then the phase space $\Phi'$ can be written in a simple form as
\begin{equation}
\label{RMDPhiMssless}
 \Phi' = {1\over m_{M_1}^2} \int 4(1 - x_{M_2})dx_{\ell_1} dx_{M_2} \approx 0.6667,
\end{equation} 
where the integration limits are $0 \le x_{\ell_1} \le 1$ and $(1-x_{\ell_1}) \le x_{M_2} \le 1$. 
It is interesting to note that the integration $\Phi'$ is finite even in the massless limit
for the final state particles, unlike the case for $\Phi$ in $\tau$ decay, due to the anti-symmetric
property of the  matrix element for the two fermions in the final state.
 
The branching ratio is then given by
\begin{eqnarray}
\label{28}
BR &=& \tau^{}_{M_1}  \Gamma =  (1 - {1\over 2} \delta_{\ell_1\ell_2})
\frac {1}{64 \pi^3} \tau_{M_1}G_F^4 |V^{CKM}_{M_1} V^{CKM}_{M_2}|^2  f_{M_1}^2 f_{M_2}^2 m_{M_1}^3 \left<m\right>_{\ell_1 \ell_2}^2 \Phi' \\ 
%BR = \frac{\tau_{M_1}}{64 \pi^3} G_F^4 |V^{CKM}_{M_1} V^{CKM}_{M_2}|^2  f_{M_1}^2 
%f_{M_2}^2 %m_{M_1}
%^3 \left<m\right>_{\ell_1 \ell_2}^2 \Phi' ,
%\label{MsonScldEqn}
% 1.42\times ...
& \approx& 10^{-29}\ |V^{CKM}_{M_1} V^{CKM}_{M_2}|^2 \left({\tau_{M_1}}
\over{1.0\times 10^{-8}\ {s}}\right)\times
\nonumber  \\
&& \left({f_{M_1}\ f_{M_2}} \over{(100\ {\mev})^2}\right)^2 
 \left({m_{M_1}}\over{1\ {\gev}}\right)^3 \left({\left<m\right>_{\ell_1 \ell_2}}\over{1\ {\ev}}\right)^2 \Phi' .
\end{eqnarray}
We have used the following constants from the PDG \cite{PDG},
the decay constant for $B$ from \cite{MILC} and for $D$ from \cite{Istvan}, for obtaining 
$\left<m\right>_{\ell_1 \ell_2}$ from the branching fractions for the various decay modes
\begin{eqnarray}
\nonumber
&& f_D = 202\ {\mev,}\  f_{Ds} = 266\  {\mev,}\ f_B = 190\  {\mev;} \\
\nonumber
&& |V_{ub}| = 0.00367,\ |V_{cd}| = 0.224,\  |V_{cs}| = 0.996;\\
\nonumber
&& \tau_K = 1.2384\times 10^{-8}\ {\s},\ \tau_D = 1.040\times 10^{-12}\ {\s},\ 
\tau_{D_s} = 4.9\times 10^{-13}\ {\s},\  \tau_B = 1.671\times 10^{-12}\ {\s};\\
\nonumber
&& m_K = 493.7\ {\mev,}\ m_D = 1869\  {\mev,}\  m_{D_s} = 1968\  {\mev,}\  m_B = 5279\ {\mev.}
\end{eqnarray}

%%%%%%%%%%%%%%%%%%%%%%%%%%%%%%%%%%%%%%%%%%%%%%%%%%%%%%.
%%%%%%%%%%%%%%%%%%%%%%%%%%%%%%%%%%%%%%%%%%%%%%%%%%%%%%
\vspace{1cm}

% A useful Journal macro
\def\jnl#1#2#3#4{{#1}{\bf #2} (#4) #3}

\def\Zphys{{\em Z.\ Phys.} }
\def\jssc{{\em J.\ Solid State Chem.\ }}
\def\jpsJ{{\em J.\ Phys.\ Soc.\ Japan }}
\def\ptps{{\em Prog.\ Theoret.\ Phys.\ Suppl.\ }}
\def\PTP{{\em Prog.\ JMKZset.\ Phys.\  }}

\def\JMP{{\em J. Math.\ Phys.} }
\def\NPB{{\em Nucl.\ Phys.} B}
\def\NP{{\em Nucl.\ Phys.} }
\def\PLB{{\em Phys.\ Lett.} B}
\def\PL{{\em Phys.\ Lett.} }
\def\PRL{\em Phys.\ Rev.\ Lett. }
\def\PRB{{\em Phys.\ Rev.} B}
\def\PRD{{\em Phys.\ Rev.} D}
\def\PRe{{\em Phys.\ Rep.} }
\def\AP{{\em Ann.\ Phys.\ (N.Y.)} }
\def\RMP{{\
em Rev.\ Mod.\ Phys.} }
\def\ZPC{{\em Z.\ Phys.} C}
\def\SCI{\em Science}
\def\CMP{\em Comm.\ Math.\ Phys. }
\def\MPLA{{\em Mod.\ Phys.\ Lett.} A}
\def\IJMPA{{\em Int.\ J.\ Mod.\ Phys.} A}
\def\IJMPB{{\em Int.\ J.\ Mod.\ Phys.} B}
\def\EPJC{{\em Eur.\ Phys.\ J.} C}
\def\PR{{\em Phys.\ Rev.} }
\def\JHEP{{\em JHEP} }
\def\cmp{{\em Com.\ Math.\ Phys.}}
\def\JPA{{\em J.\  Phys.} A}
\def\CQG{\em Class.\ Quant.\ Grav. }
\def\ATMP{{\em Adv.\ Theoret.\ Math.\ Phys.} }
\def\ibid{{\em ibid.} }


\newpage

\begin{thebibliography}{99}
\small
\baselineskip=14pt

%%%%%%

\bibitem{review}
For reviews on flavor physics, see {\it e.g.}, 
M. Artuso, B. Gavela, B. Kayser, C. McGrew, P. Rankin, and E.D. Zimmerman, 
FERMILAB-CONF-01-428, in eConf C010630:P2001 (2001);
A. Masiero, S.K. Vempati, and O. Vives, New J. Phys. {\bf 6}, 202 (2004) [{\tt arXiv:hep-ph/0407325}].

\bibitem{BargerReview}
For a recent review, see {\it e.g.},
V. Barger, D. Marfatia and K. Whisnant,  Int.~J.~Mod.~Phys.~{\bf E12}, 569-647 (2003) 
[{\tt arXiv:hep-ph/0308123}]; for earlier comprehensive discussions of neutrino physics see 
{\it e.g.}, {\it Massive Neutrinos in Physics and Astrophysics} by R. N. Mohapatra and P. B. 
Pal (World Scientific 2004); {\it Physics of Neutrinos and Applications to Astrophysics} by M. Fukugita and T. Yanagida (Springer-Verlag 2003); {\it The Physics of Massive Neutrinos} by B. Kayser, F. Gibrat-Debu and F. Perrier (World Scientific 1989).

\bibitem{seesaw}
 P. Minkowski, Phys. Lett. {\bf B67}, 421 (1977); T. Yanagida, in {\it Proc. of the Workshop on Grand Unified Theory and Baryon Number of the Universe}, KEK, Japan, 1979; M. Gell-Mann, P. Ramond  and R. Slansky in {\it Sanibel Symposium}, February 1979, CALT-68-709 [{\tt retroprint arXiv:hep-ph/9809459}], and in {\it Supergravity}, eds. D. Freedman {\it et al}. (North Holland, Amsterdam, 1979); S.L. Glashow in {\it Quarks and Leptons, Cargese,} eds. M. Levy {\it et al.} (Plenum, 1980, New York), p. 707; R. N. Mohapatra and G. Senjanovic,  Phys. Rev. Lett. {\bf 44}, 912 (1980).

\bibitem{RparitySUSY}
C. S. Aulakh and R. N. Mohapatra,  Phys. Lett. {\bf B119}, 136 (1982); L. J. Hall and M. Suzuki, Nucl. Phys {\bf B231}, 419 (1984); G. G. Ross and J. W. F. Valle, Phys. Lett. {\bf B151}, 375 (1985); J. Ellis, G. Gelmini, C. Jarlskog, G. G. Ross and J. W. F. Valle, Phys. Lett. {\bf B150}, 142 (1985); S. Dawson, Nucl. Phys. {\bf B261}, 297 (1985); M. Drees, S. Pakvasa, X. Tata and T. ter. Veldhuis, Phys. Rev. {\bf D57}, R5335 (1998) [{\tt arXiv:hep-ph/9712392}]; E. J. Chun, S. K. Kang, C. W. Kim and U. W. Lee, Nucl. Phys. {\bf B544}, 89 (1999)  
[{\tt arXiv:hep-ph/9807327}]; V. Barger, T. Han, S. Hesselbach and D. Marfatia, Phys. Lett. {\bf B538}, 346 (2002) [{\tt arXiv:hep-ph/0108261}]; for a recent review see R. Barbieri {\it et al}, {\tt arXiv:hep-ph/0406039}.

\bibitem{LRModels}
J. C. Pati and A. Salam, Phys. Rev. {\bf D10}, 275 (1974); R. N. Mohapatra and J. C. Pati, Phys. Rev {\bf D11}, 566, 2558 (1975); G. Senjanovic and R. N. Mohapatra, Phys. Rev. {\bf D12}, 1502 (1975).

\bibitem{SO10SUSYGUT}
J. A. Harvey, P. Ramond and D. B. Reiss, Nucl. Phys. {\bf B199}, 223 (1982); S. Dimopoulos, L. J. Hall and S. Raby, Phys. Rev. Lett. {\bf 68}, 1984 (1992); L. J. Hall and S. Raby, Phys. Rev {\bf D51}, 6524 (1995) [{\tt arXiv:hep-ph/9501298}].

\bibitem{ZeeModel}
A. Zee, Phys. Lett. {\bf B93}, 389 (1980) [{\it Erratum - ibid.} {\bf B95}, 461 (1980)]; Phys. Lett {\bf B161}, 141 (1985).

\bibitem{MaModels}
E. Ma and U. Sarkar, Phys. Rev. Lett. {\bf 80}, 5716 (1998); E. Ma and G. Rajasekaran, Phys. Rev. {\bf D64}, 113012 (2001) [{\tt arXiv:hep-ph/0106291}]; E. Ma, Mod. Phys. Lett. {\bf A17}, 289 (2002)
 [{\tt arXiv:hep-ph/0201225}]; K. S. Babu, E. Ma and J. W. Valle, Phys. Lett. {\bf B552}, 207 (2003)
  [{\tt arXiv:hep-ph/0206292}]; E. Ma, Mod. Phys. Lett. {\bf A17}, 2361 (2002)
   [{\tt arXiv:hep-ph/0211393}]. 

\bibitem{ExtraDim}
Nima Arkani-Hamed, Savas Dimopoulos, Gia Dvali and John March-Russell, Phys. Rev. {\bf D65}, 024032 (2002) [{\tt arXiv:hep-ph/9811448}]; Keith R. Dienes and Ina Sarcevic, Phys. Lett. {\bf B500}, 133 (2001) [{\tt arXiv:hep-ph/0008144}].

\bibitem{babuleung}
 K.S. Babu and C.N. Leung, Nucl.~Phys.~{\bf B619}, 667 (2001)  [{\tt arXiv:hep-ph/0106054}].
  
\bibitem{nuless}
W. H. Furry, Phys. Rev. {\bf 56}, 1184 (1939); for early reviews see, Primakoff and Rosen, Rep. Prog. Phys. {\bf 22}, 121 (1959); Ann. Rev. Nucl. Part. Sci. {\bf 31}, 145 (1981).

\bibitem{ElliottEngel}
For recent review see {\it eg.} S.R. Elliott and J. Engel, J. Physics. {\bf G30} R183 (2004)
 [{\tt arXiv:hep-ph/0405078}].

\bibitem{DoiKotani}
M. Doi, T. Kotani and E. Takasugi, Prog. Theor. Phys. Suppl. {\bf 83}, 1 (1985).

\bibitem{TauDecay}
A. Ilakovac, B. A. Kniehl and A. Pilaftsis, Phys. Rev {\bf D52}, 3993 (1995); A. Ilakovac and A. Pilaftsis, Nucl. Phys. {\bf B437}, 491 (1995); A. Ilakovac, Phys. Rev. {\bf D54}, 5653 (1996)
 [{\tt arXiv:hep-ph/9608218}].

\bibitem{Barbero}
 L.S. Littenberg and R.E. Shrock, Phys. Rev. {\bf D46}, R892 (1992); C. Barbero, G. Lopez Castro and A. Mariano, Phys. Lett. {\bf B566}, 98 (2003) [{\tt arXiv:nucl-th/0212083}].

\bibitem{CSLim}
C.S. Lim, E. Takasugi and M. Yoshimura, {\tt arXiv:hep-ph/0411139}.

\bibitem{TRizzo}
T. G. Rizzo, Phys. Lett. {\bf B116}, 23 (1982); C. A. Heusch and P. Minkowski, Nucl. Phys. {\bf B416}, 
3 (1994).

\bibitem{Rodejohann}
M. Flanz, W. Rodejohann and K. Zuber, Phys. Lett. {\bf B473}, 324 (2000), {\it Erratum - ibid.} {\bf B480}, 418 (2000)  [{\tt arXiv:hep-ph/9911298}].

\bibitem{Flanz}
M. Flanz, W. Rodejohann and K. Zuber, Eur. Phys. J. {\bf C16}, 453 (2000) [{\tt arXiv:hep-ph/9907203}]; W. Rodejohann and K. Zuber, Phys. Rev. {\bf D63}, 054031 (2001)  [{\tt arXiv:hep-ph/0011050}].

\bibitem{AliBorisov}
A. Ali, A.V. Borisov and N.B. Zamorin, Eur. Phys. J. {\bf C21}, 123 (2001) 
[{\tt arXiv:hep-ph/0104123}].



\bibitem{MNSP}
Z. Maki, M. Nakagawa and S. Sakata, Prog. Theor. Phys. {\bf 28}, 870 (1962);
B. Pontecorvo, Sov. Phys. JETP  {\bf 26}, 984 (1968).

\bibitem{Smirnov}
Michele Frigerio and Alexei Yu. Smirnov, 
Nucl. Phys. {\bf B640}, 233 (2002)  [{\tt arXiv:hep-ph/0202247}].

\bibitem{SuperK}
M. Ishitsuka [Super-Kamiokande Collaboration], {\tt arXiv:hep-ex/0406076}.

\bibitem{Kearns}
Super-Kamiokande Collaboration, Y. Ashie {\it et al.}, Phys. Rev. Lett. {\bf 93}, 101801 (2004) 
[{\tt arXiv:hep-ex/0404034}].

\bibitem{SKNew}
Super-Kamiokande Collaboration, Y. Ashie {\it et al.}, {\tt arXiv:hep-ex/0501064}, 
submitted to Phys. Rev. D.

\bibitem{CHOOZ}
M. Apollonio {\it et al.}, Phys. Lett. {\bf B466}, 415 (1999); T. Nakaya [Super-Kamiokande Collaboration], eConf {\bf  C020620}, SAAT01 (2002) [{\tt arXiv:hep-ex/0209036}].

\bibitem{Bahcall}
J. N. Bahcall, M. C. Gonzlez-Garcia, Carlos Pe\~{n}a-Garay, JHEP 0408 (2004) 016
 [{\tt arXiv:hep-ex/0406294}].

\bibitem{Seljak}
U. Seljak {\it et al.}, {\tt arXiv:astro-ph/0407372}, submitted to PRD.

\bibitem{Adam}
Vernon Barger, Danny Marfatia and Adam Tregre, Phys. Lett. {\bf B595}, 55 (2004)
 [{\tt arXiv:hep-ph/0312065}].

\bibitem{OtherSeljak}
U. Seljak {\it et al.}, Phys. Rev. {\bf D71}, 043511 (2005) [{\tt arXiv:astro-ph/0406594}].

\bibitem{Glashow}
V. Barger, S.L. Glashow, D. Marfatia and K. Whisnant, Phys. Lett. {\bf B532}, 15 (2002)
 [{\tt arXiv:hep-ph/0201262};
 S. Pascoli and S.T.Petcov, Phys. Lett. {\bf B544}, 239 (2002)
 [{\tt arXiv:hep-ph/0205022}].

\bibitem{Pas}
H. V. Klapdor-Kleingrothaus, H.P\"{a}s and A. Yu. Smirnov, Phys. Rev. {\bf D63}, 073005 (2001).

\bibitem{Fogli}
G. L. Fogli, E. Lisi, A. Marrone, A. Melchiorri, A. Palazzo, P. Serra, J. Silk, Phys. Rev. {\bf D70}, 113003 (2004) [{\tt arXiv:hep-ph/0408045}].

\bibitem{WR}
 W.~Rodejohann, J.~Phys.~{\bf G 28}, 1477 (2002). 
 
\bibitem{Klapdor}
H.V. Klapdor-Kleingrothaus, I.V. Krivosheina, A. Dietz and O. Chkvorets, Phys. Lett. {\bf B586}, 
198 (2004) [{\tt arXiv:hep-ph/0404088}].

\bibitem{CLEO}
CLEO Collaboration, D. Bliss {\it et al.}, Phys. Rev. {\bf D57}, 5903 (1998) 
[{\tt arXiv:hep-ex/9712010}].

\bibitem{Littenberg}
L.S. Littenberg and R.E. Shrock, Phys. Rev. Lett. {\bf 68}, 443 (1992); 
C. Dib, V. Gribanov, S. Kovalenko and I. Schmidt, Phys. Lett. {\bf B493}, 82 (2000)
[{\tt arXiv:hep-ph/0006277}].

\bibitem{PDG}
PDG, S. Eidelman {\it et al.} Phys. Lett. {\bf B592}, 1 (2004).

\bibitem{Grossman}
Y. Grossman, Z. Ligeti and E. Nardi, Phys. Rev. {\bf D55}, 2768 (1997) 
[{\tt arXiv:hep-ph/9607473}]. 

\bibitem{SINDRUM}
SINDRUM II Collaboration, J. Kaulard {\it et al.}, Phys. Lett. {\bf B422}, 334 (1998).

\bibitem{Zuber}
K. Zuber, {\tt arXiv:hep-ph/0008080}.

\bibitem{Simkovic}
P. Domin, A. Faessler, S. Kovalenko and F. Simkovic, Phys. Rev. {\bf C70}, 065501 (2004) [{\tt arXiv:nucl-th/0409033}].

\bibitem{Missimer}
J.H. Missimer, R.N. Mohapatra and Nimai C. Mukhopadhyay, Phys. Rev. {\bf D50}, 2067 (1994).

\bibitem{Faessler}
F. Simkovic, A. Faessler,  S. Kovalenko and P. Domin, Phys. Rev. {\bf D66}, 033005 (2002) 
[{\tt arXiv:hep-ph/0112271}].

\bibitem{Takasugi}
E. Takasugi, Nucl. Instrum. Meth. {\bf A503}, 252 (2003).

\bibitem{Aoki}
M. Aoki, Nucl. Instrum. Meth. {\bf A503}, 258 (2003).

\bibitem{WayneHu}
W. Hu, D. J. Eisenstein and M. Tegmark, Phys. Rev. Lett. {\bf 80}, 5255 (1998). 

\bibitem{Ivanov}
Mikhail A. Ivanov and Sergey G. Kovalenko, Phys. Rev. {\bf D71}, 053004 (2005)  [{\tt arXiv:hep-ph/0412198}].

\bibitem{collider}
V.~Barger and R.~Phillips, {\it Collider Physics},  Addison-Wesley (1987).

\bibitem{MILC}
MILC Collaboration, C. Bernard {\it et al.}, Phys. Rev. {\bf D66}, 094501 (2002)
 [{\tt arXiv:hep-lat/0206016}].

\bibitem{Istvan}
Istvan Danko for the CLEO Collaboration, {\tt arXiv:hep-ex/0501046}.





\end{thebibliography}
\end{document}